\documentstyle[prl,aps,multicol,graphicx]{revtex}

\graphicspath{{figures/}}

\begin{document}

\tighten
\draft
\title{Low temperature acoustic properties of amorphous silica and the Tunneling Model}

\author{EunJoo Thompson, G. Lawes, J. M. Parpia, R. O. Pohl$^{*}$}
\address{Laboratory of Atomic and Solid State Physics, Cornell University, 
Ithaca, New York  14853-2501}
\date{\today} 
\maketitle

\begin{abstract}
Internal friction and speed of sound of {\em a}-SiO$_{2}$ was measured
above 6\,mK using a torsional oscillator at 90\,kHz, controlling for
thermal decoupling, non-linear effects, and clamping losses.  Strain
amplitudes $\epsilon_{\rm A}$\,$=$\,$10^{-8}$ mark the transition
between the linear and non-linear regime.  In the linear regime, excellent
agreement with the Tunneling Model was observed for both the internal
friction and speed of sound, with a cut-off energy of $\Delta_{\rm o,
min}/{\rm k}_{\rm B}$\,=\,6.6\,mK. In the non-linear regime, two
different behaviors were observed.  Above 10\,mK the behavior was
typical for non-linear harmonic oscillators, while below 10\,mK a
different behavior was found.  Its origin is not understood.
\end{abstract}


\begin{multicols}{2}

\narrowtext

The low temperature acoustic, thermal, and dielectric properties of
amorphous solids have long been successfully described by the
phenomenological Tunneling Model (TM).  In this model, the low energy
localized vibrational excitations, a common feature of amorphous
solids, are described by non-interacting two-level defects which are
thought to be caused by tunneling of atoms or groups of atoms between
nearly degenerate potential minima.  The excitation energy between the
two lowest states of the double well potential, E =
$\sqrt{\Delta^{2}+\Delta_{\rm{o}}^{2}}$, is determined by the
asymmetry, $\Delta$ and the tunneling splitting,
$\Delta_{\rm{o}}$\cite {phillips87}.  Low temperature internal
friction and speed of sound measurements on {\em a}-SiO$_{2}$ using
the torsional oscillator technique between 66 and 160\,kHz above
50\,mK have shown excellent agreement with the TM
\cite{vancleve91,toppPRB99}.  However, several acoustic and dielectric
experiments have indicated deviations from this model below 100\,mK
\cite{Nishiyama92,esquinazi92,classen94,Rogge97,classen99} and have
been interpreted as evidence for tunneling defect
interactions\cite{enss97,burin98}.  This discrepancy provided the
impetus for the present study in which we extended the acoustic
measurements to 6\,mK.

In low temperature acoustic measurements, a major cause of uncertainty
are thermal decoupling caused by spurious heat input and self heating,
non-linear effects resulting from moderate strain amplitudes, and lack
of knowledge of the influence of
mounting\cite{vancleve91,esquinazi92,classen94,classen99,burin98}. 
Taking particular care to control these problems, we report here
measurements in the extended temperature range which are in excellent
agreement with the predictions of the TM with a low energy cut-off in
the tunneling state spectrum, $\Delta_{\rm o,min}/{\rm k}_{\rm B}$ =
6.6\,mK. These results emphasize the extreme care with which low
temperature acoustic work must be carried out.

The amorphous silica sample, Suprasil - W, ($<$ \,5\,ppm
OH$^{-}$\,impurities, 4\,mm diameter, 22.33\,mm long) was mounted in a
torsional composite oscillator resonating at $\sim$\,90\,kHz
\cite{cahilljvc89} in a dilution refrigerator (0.07 - 2K) and in a
vibrationally isolated dilution refrigerator with a demagnetization
stage (0.006 - 0.100\,K)\cite{parpia85}.  In the latter, the sample
was surrounded by a Nb tube (5.4\,mm i.d., 60\,mm long) in order to
shield it from residual magnetic fields except for the earth's field
($<$\,0.5\,G).

Before presenting our results on the acoustic properties, we describe
the thermal and strain studies that are crucial for avoiding
experimental errors.  We begin by considering {\it{thermal
decoupling}} of the sample.  The temperature of the sample was
measured directly by epoxying a 1k$\Omega$ RuO$_{2}$ Dale resistor
onto the free end of the undriven {\em a}-SiO$_{2}$ sample, with the
leads (76\,$\mu$m diameter Evanohm) thermally anchored along its
length (see Fig.\,1 inset).  A twin Dale resistor was attached to the
base.  RuO$_{2}$ thick film resistors have been successfully used as
thermometers from 0.015 to 80\,K in magnetic fields up to
20\,T\cite{li86}.  The resistances for the two resistors, read with a 
self-balancing resistance bridge (Linear Research, LR700) using a 
power\,=\,10$^{-15}$W and calibrated
against a He$^{3}$ melting curve thermometer, were found to have
identical temperature dependencies above 10\,mK with very little
scatter, see Fig.\,1a.  Below 10\,mK, however, irreproducible
resistances for both resistors were observed (error bars).  The reason
for this irreproducibility is not known, nor are we aware of any
measurements on these resistors below 15\,mK. Because of this
irreproducibility, thermal decoupling of the undriven sample can be
definitely ruled out only above 10\,mK. Heating of the oscillator
through vibrational noise in this cryostat below 10\,mK is nonetheless
considered to be unlikely.  The lowest frequency mode of the sample, 
f\,$\approx$\,3\,kHz, is that in which the stiff oscillator bends the thin Be:Cu
torsion rod (1\,mm diameter, 3\,mm long).  Because of its relatively
large metallic thermal conductivity, this heat will be easily removed
through the base.  In addition, a torsional oscillator in the same
refrigerator connected by a hollow Be:Cu torsion rod previously showed
no signs of thermal decoupling to 1\,mK\cite{Parpia83}.

Heating can also occur when the oscillator is driven.  The strain
amplitude, $\epsilon_{\rm A}$, is defined as the maximum angular
displacement between the two ends of the silica sample.  The
electrical measurement of $\epsilon_{\rm A}$ was calibrated by
reflecting a laser beam off the sample and measuring its deflection
with a photodiode\cite{liu99}.  The power loss from mechanical
dissipation is $Q^{-1}f_{\rm{o}}\frac{1}{2}GV\epsilon_{\rm{A}}^{2}$,
where $Q^{-1}$ is the internal friction, $f_{\rm{o}}$ the resonant
frequency, $G$ the shear modulus, and $V$ the volume of the
oscillator.  The thermal resistance, R$_{\rm{th}}$, of the oscillator
was measured by heating the sample thermometer (inset of Fig.\,1) with
a controlled power input from the LR700.  The data are close to the
dashed line which was calculated ignoring the thermal resistance of
the glass sample (see Fig.\,1 caption).  The explanation may be that
heat enters the glass over its entire surface by way of the resistor
leads.  R$_{\rm{th}}$ was calculated by adding the thermal resistance
of the glass to that shown as a dashed line, and was used to determine
the upper limit of the temperature rise of the driven
oscillator for a measured mechanical power loss.  For example, at
10\,mK, $\epsilon_{\rm{A}}\,=\,10^{-8}$ would result in $\Delta$T =
40$\mu$K. It should be noted, however, that $\epsilon_{\rm{A}} \sim
10^{-7}$ would result in a temperature rise of 4\,mK at 10\,mK in our
amorphous sample, a significant source of self heating and error.

\begin{figure}[p] 
\includegraphics[scale=0.45]{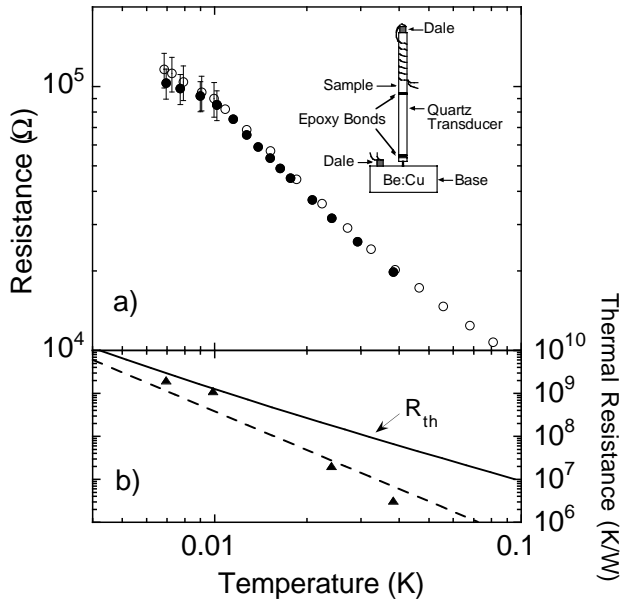}
\vspace{-.6in} \caption{a) Resistance versus temperature for
1\,k$\Omega$ RuO$_{2}$ Dale thick film resistors mounted on the sample
(solid circles) and the base (open circles).  The inset shows the
experimental setup for this experiment.  b) Measured (triangles) and
calculated thermal resistances of the torsional oscillator.  The solid
line is the calculated thermal resistance,
R$_{\rm{th}}$\,=\,$\frac{87500}{T^{2}}$
(K$^{3}$/W)\,+\,$\frac{181}{T^{3}}$(K$^{4}$/W)\,+\,$\frac{204}{T^{3}}$(K$^{4}$/W),
using the method outlined in Ref.\,\protect\onlinecite{swartz89} and
the data table of Ref.\,\protect\onlinecite{cheeke76}, where the terms
are the thermal resistances of {\em a}-SiO$_{2}$, Casimir, and
boundary respectively.  The dashed line is the thermal resistance
without {\em a}-SiO$_{2}$.  (The thermal resistance of the wires
connecting the sample to the cryostat was much larger than that of the
mounted oscillator.)}
\end{figure} 

The elastic properties of the {\it{background}} were measured by replacing
the amorphous sample with a crystal quartz sample of equal size which
has negligible internal dissipation.  The frequency traces of the
background oscillator (quartz on quartz) at different strain
amplitudes showed a Lorentzian line shape at all amplitudes and
temperatures.  Fig.\,2a shows two traces taken at nominally
2\,mK (i.e. ignoring the $\Delta$T due to self-heating).  They show
that neither the epoxy, nor the Be:Cu metal base lead to non-linear
behavior even when the strain amplitude was varied by a factor of 100. 
The background internal friction (clamping losses) was found to be
slightly temperature dependent while the speed of sound was
independent of temperature to within $\sim$\,0.1\,ppm, as shown
in Fig.\,3.

Elastic measurements are sensitive to {\it{non-linear effects}} as can
be seen in frequency traces of a driven {\em a}-SiO$_{2}$ at
increasing driving voltages (peak power dissipation).  These traces
are plotted in Fig.\,2b-d at 27\,mK, 10.6\,mK and 7\,mK respectively. 
The dashed curve is the normalized response of the oscillator carrying
a crystal quartz sample.  Above 10\,mK, the lineshapes for {\em
a}-SiO$_{2}$ are clearly resolved from that of the quartz sample
(background) as seen in Fig.\,2b.  In contrast to the background, the
frequency response of the oscillator with the amorphous sample shows a
variety of non-linear behaviors.  Above 10\,mK, with increasing peak
strain amplitude, the oscillator exhibits behavior typical for a
non-linear harmonic oscillator.  (This non-linearity is not seen above
50\,mK where it is masked by self-heating\cite{liu99}.)  Below 10\,mK,
however, (see Fig.\,2c and 2d) the non-linear behavior differs
dramatically.  With increasing strain amplitudes, the frequency
response does not lean to higher frequencies.  Instead the response
has jumps, plateaus and oscillations.  Since the background oscillator
shows no such behavior, we conclude that the tunneling entities in
{\em a}-SiO$_{2}$ are responsible.  This is the first evidence that
the tunneling entities themselves may change their behavior below
10\,mK, although we do not understand the nature of the tunneling
entities that can bring about these non-linearities.

The internal friction was only determined when the frequency
dependence of the oscillator was clearly in the linear regime, i.e.
fit a Lorentzian curve well (solid curves in Fig.\,2).  In the linear
regime, the internal friction was found to be independent of the
driving power and $\Delta$T to be negligible.  At larger strain
amplitudes (in the non-linear regime), the fits yielded higher
internal frictions and speeds of sound.  As an example, consider
Fig\,2c; for a peak $\epsilon_{\rm A}$\,=\,1.7\,$\times$\,10$^{-8}$
and a power\,=\, 3\,$\times$\,10$^{-14}$\,W, using a half-width
of the non-Lorentzian would lead to an erroneous increase of the
internal friction by a factor of 2 and an increase of the speed of
sound by $\sim$1\,ppm.  Note that in the most recent investigation of
{\em a}-SiO$_{2}$\cite{classen99}, a double paddle oscillator
operating at $\epsilon_{\rm A}$ $\sim$10$^{-7}$ was used, which should
lead to considerable non-linearities at low temperatures, at least in
our geometry.

\begin{figure}[p] 
\vspace{-1.5in} 
\includegraphics[scale=0.45]{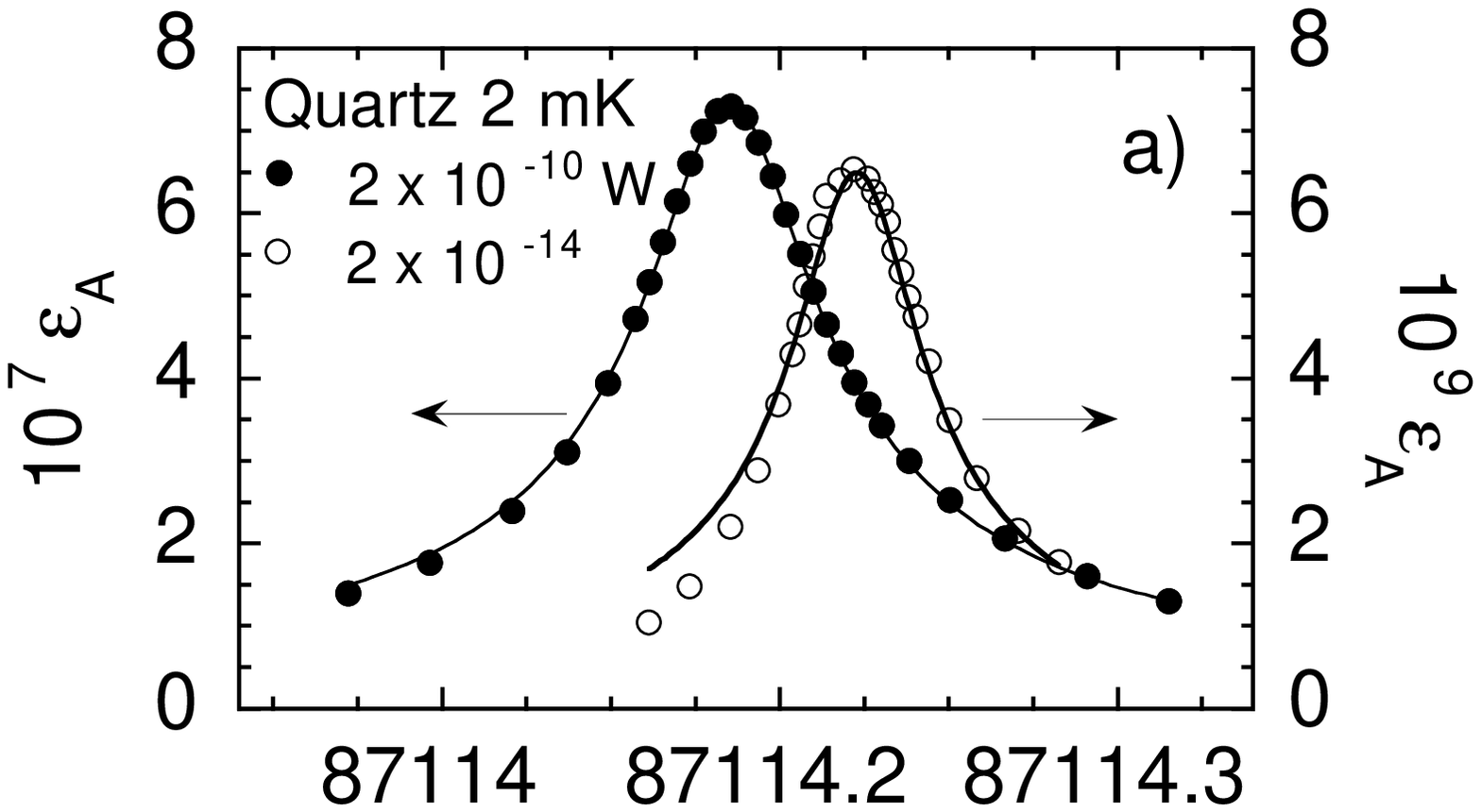} \newline
\includegraphics[scale=0.45]{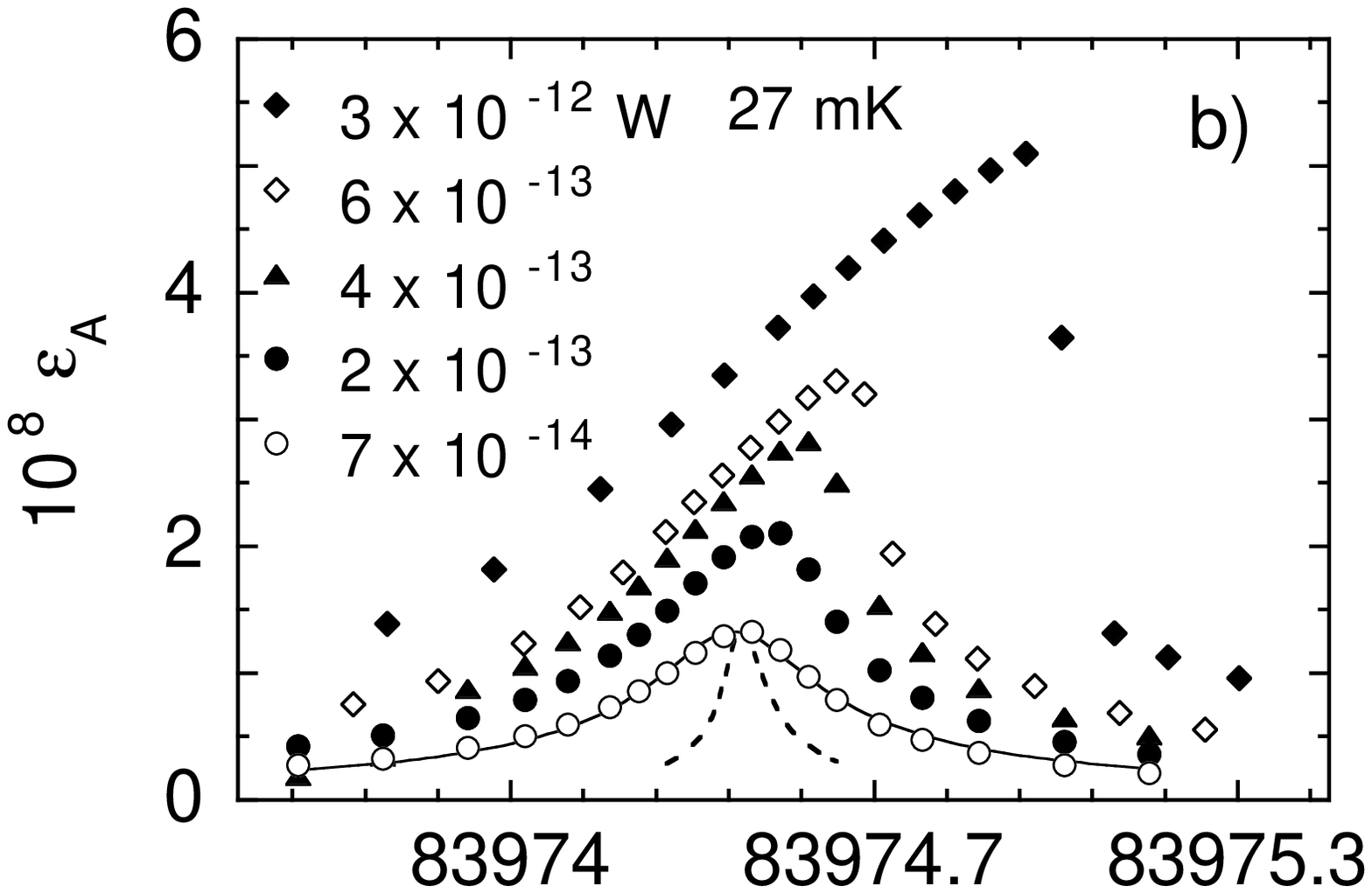} \newline
\includegraphics[scale=0.45]{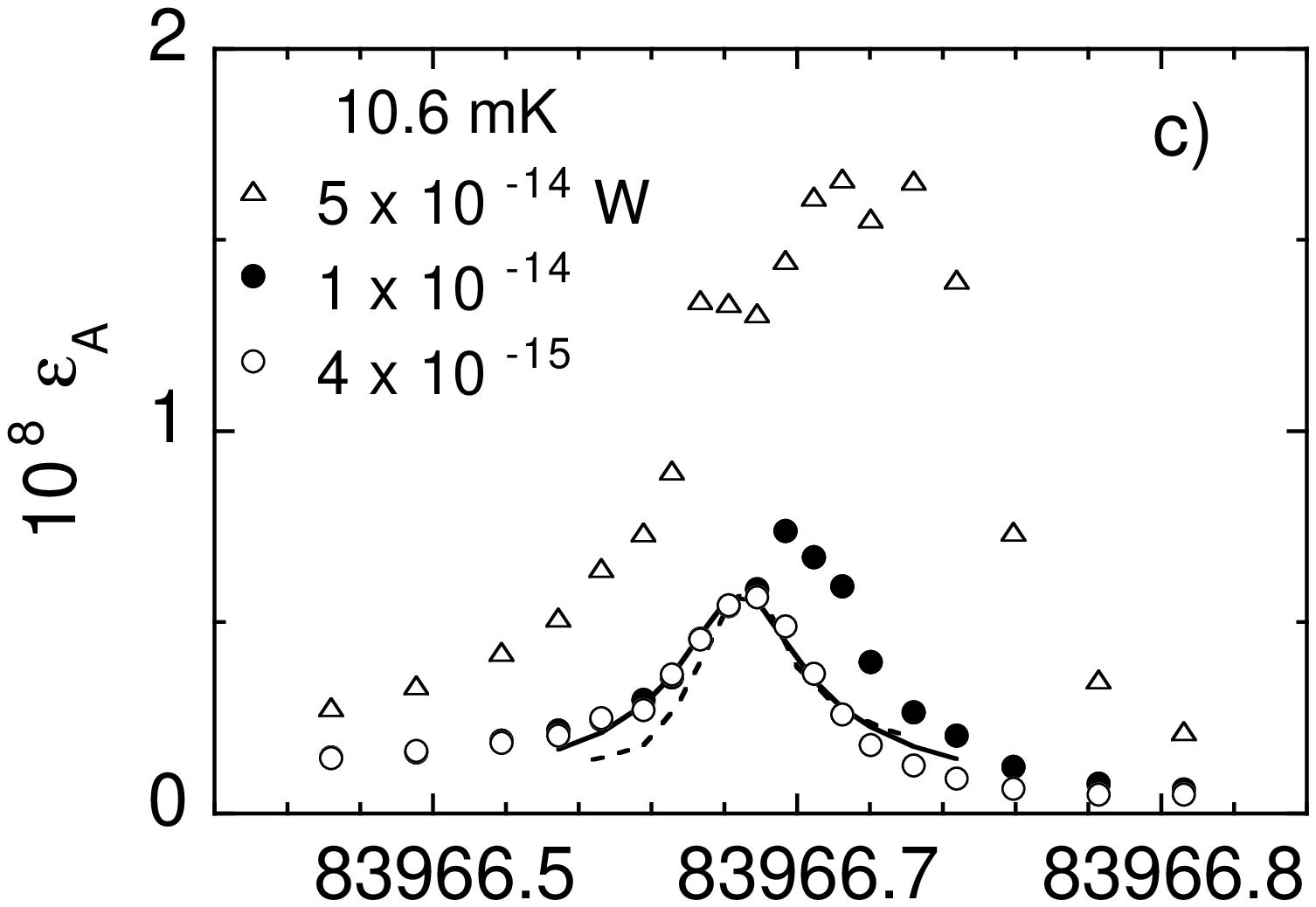} \newline
\includegraphics[scale=0.45]{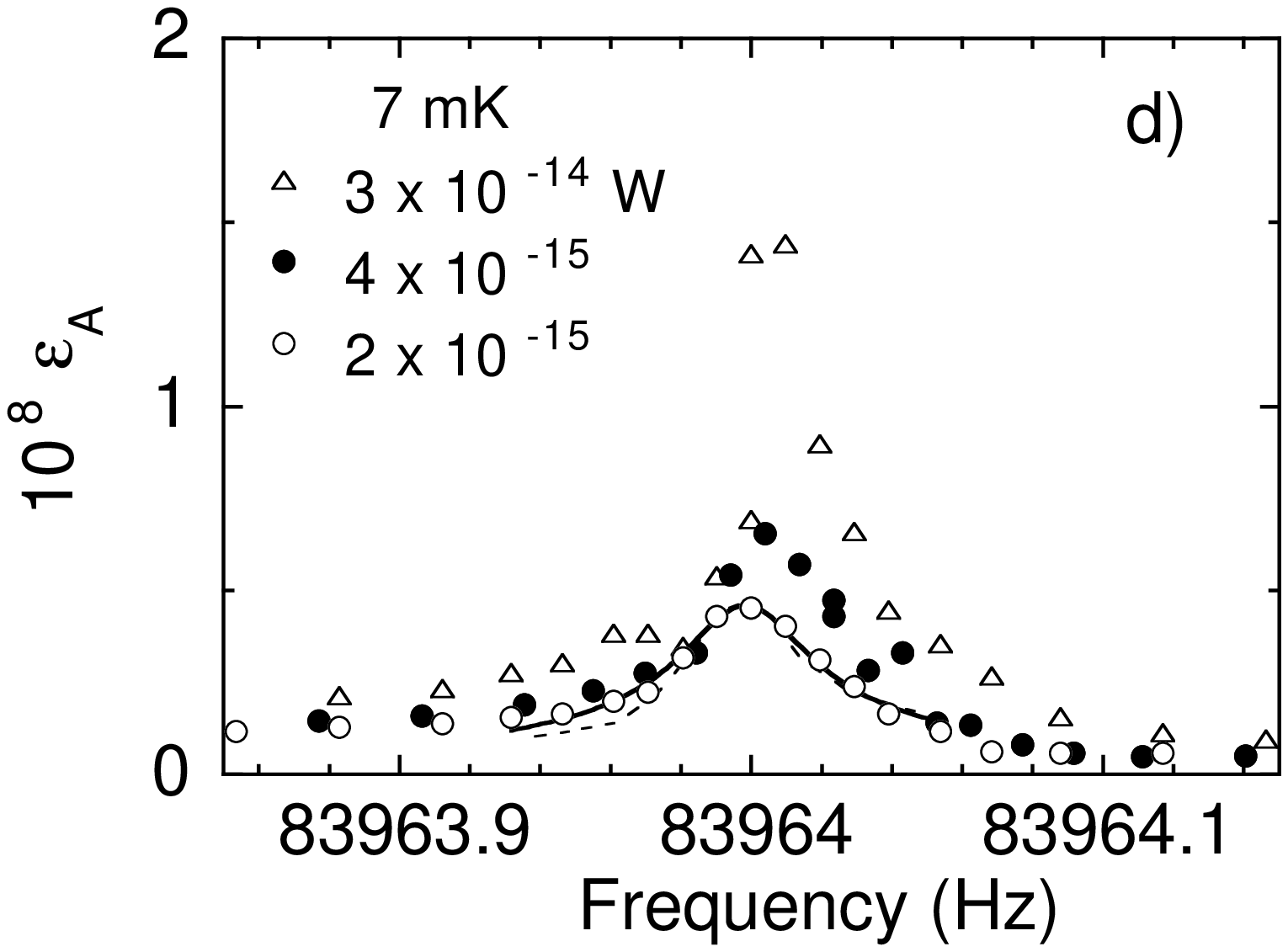} \newline
\caption{Frequency response of the background (quartz) and of {\em
a}-SiO$_{2}$.  a) Lorentzian line shapes of quartz measured at 
$\sim$\,2\,mK
leading to a peak power dissipation of 2\,$\times$\,$10^{-14}$\,W and 
2\,$\times$\,$10^{-10}$\,W are plotted
using the right and left y-axes respectively.  Solid curves are
Lorentzian fits to the data.  b-d) frequency response of {\em a}-SiO$_{2}$ at
27\,mK, 10.6\,mK, and 7\,mK for different peak powers.  Dashed curves are
normalized background Lorentzians at each temperature.}
\end{figure} 

\begin{figure}[p] 
\includegraphics[scale=0.55]{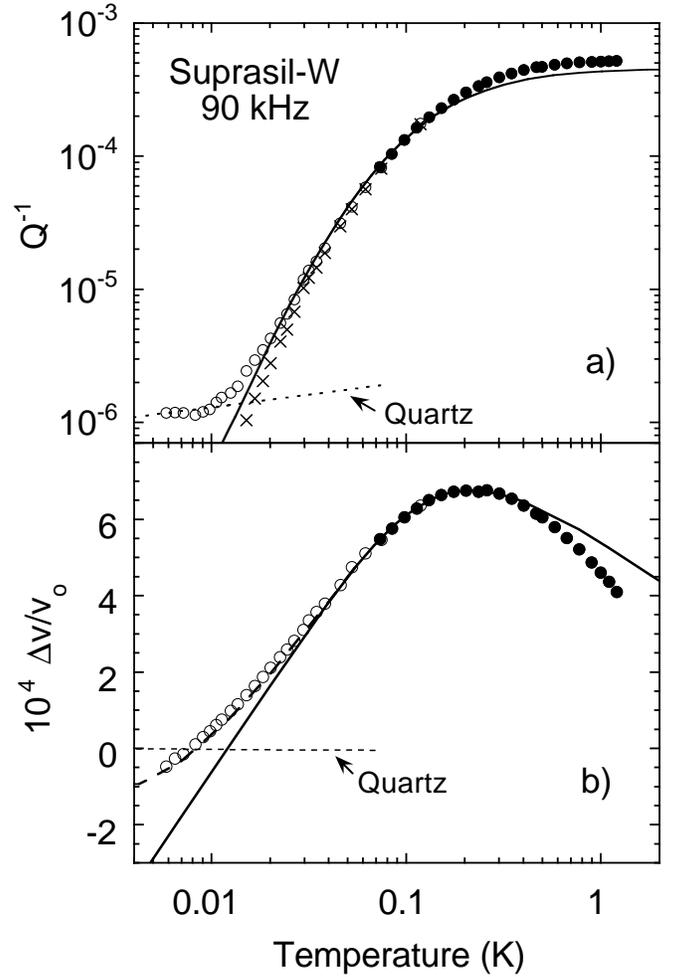} 
\caption{a) Internal friction and b) relative change of the transverse
speed of sound of {\em a}-SiO$_{2}$ (Suprasil-W) at 90\,kHz and the
background.  Solid circles, in dilution cryostat; open circles, in
demagnetization cryostat; x's, after subtraction of the background;
dashed lines, quartz sample (background); solid curves, TM prediction
with parameters from sound velocity measurements; long dash curve, TM
prediction with a cut-off energy of $\Delta_{\rm o, min}/{\rm k}_{\rm
B}$\,=\,6.6\,mK.}
\end{figure}

With the previously identified experimental problems avoided, Fig.\,3
shows the internal friction, Q$^{-1}$, and the relative change of the
transverse speed of sound, $\frac{\Delta\rm{v}}{\rm{v}_{o}}$ (see
eq.\,1) of {\em a}-SiO$_{2}$ which were determined from the Lorentzian
frequency responses, as shown in Fig.\,2 as solid curves.  Sample
heating was always less than 0.5\% of the temperature.  The solid and
the open circles show the excellent agreement of the data from the two
cryostats.  Below 10\,mK, the internal friction approaches a nearly
temperature independent value very close to that measured on the
quartz sample (dashed line).  This close agreement is also evidenced
in Figs.\,2c and 2d with the frequency responses obtained on both
samples (solid and dashed Lorentzians).  We conclude that the internal
friction measured on the {\em a}-SiO$_{2}$ sample below 10\,mK is
dominated by the background, and use the dashed line to derive the
internal friction of the {\em a}-SiO$_{2}$ without this background,
shown as x's in Fig.\,3a, following the method outlined in
Ref\,\onlinecite{toppPRB99}.  No such correction needs to be applied
to the speed of sound, because the background frequency shift is
independent of temperature (dashed line in Fig.\,3b).  The solid
curves in Fig.\,3 are fits to the TM using a tunneling strength
C\,=\,3.1\,$\times$\,10$^{-4}$ and a crossover temperature 
T$_{\rm{co}}$\,=\,0.08\,K, both identical to our previous published values based on
the measurements above 50\,mK\cite{toppPRB99}.  The internal friction
shows no deviation from the TM prediction (below 15\,mK, separating
background from the internal friction by the tunneling states would
involve large errors).  The speed of sound also agrees with the TM
prediction between 25 and 500\,mK. The deviation of $\frac{\Delta
\rm{v}}{\rm{v_{o}}}$ for T\,$>$\,500\,mK is experimentally well
established\cite{enss97}, indicating the existence of channels for
defect relaxation other than by single phonon emission.  Very
remarkable and new, however, is the deviation of 
$\frac{\Delta\rm{v}}{\rm{v_{o}}}$ from the logarithmic temperature dependence below
25\,mK which is unambiguous at least to 10\,mK (the uncertainty below
10\,mK is only due to the thermometer calibration as discussed earlier).  This
deviation is consistent within the standard TM assuming a cut-off
$\Delta_{\rm o, min}/{\rm k}_{\rm B}$ of the energy distribution of
the tunneling states, which affects the speed of sound
v\cite{vancleve91}:
\begin{equation}
\frac{\rm{v}(\rm{T}) - \rm{v_{o}}}{\rm{v_{o}}} = \frac{\Delta 
\rm{v}}{\rm{v_{o}}} = 
\rm{C(\ln}\frac{\rm{T}}{\rm{T_{\rm{o}}}} +
\frac{\Delta_{\rm{o,min}}/\rm{k_{\rm{B}}}}{2\rm{T}}), 
\label{eq:Emin} 
\end{equation}
where v$_{\rm{o}}$ is the speed of sound at some reference temperature
$\rm{T}_{\rm{o}}$, and C is the tunneling strength.  The long-dashed
curve, calculated with $\Delta_{\rm o, min}/{\rm k}_{\rm B}$ = 6.6 mK,
fits the data in Fig.\,3b well.  Note that a cut-off should not affect
the internal friction significantly, in agreement with the
experimental findings.  Evidence for a cut-off energy has also been
obtained recently from heat pulse measurements of {\em a}-SiO$_{2}$ ($\Delta_{\rm o,
min}/{\rm k}_{\rm B}$\,=\,3.1\,mK)\cite
{strehlow99physB}, and from dielectric measurements on a
multi-component alumosilicate glass (12.2\,mK)\cite{strehlow99short}. 
While such a cut-off may indeed be caused by interaction between the
tunneling defects, our data show no evidence for the effects of such
interactions beyond this gap.

In conclusion, our measurements have shown no evidence below 500\,mK
for a deviation from the predictions of the standard Tunneling Model
provided we include a cut-off energy in the mK range.  Our results disagree with the
earlier studies\cite{esquinazi92,classen94,classen99,burin98} in which
much weaker temperature dependences of both internal friction and
speed of sound than predicted by the model had been observed, although at
different frequencies ($\leq$\,14\,kHz).  Those measurements had been
interpreted as evidence for defect interactions at temperatures as
high as 100\,mK\cite{enss97,burin98}.  We
suggest that an alternative explanation may be based, at least in 
part, on
experimental problems common to acoustic experiments such as
self-heating and non-linear responses at large strain amplitudes,
which can lead to significant errors as we have shown here.  It
appears essential that all previous evidence for defect interactions
be inspected meticulously in order to convincingly exclude these 
sources of errors.  This is not feasible with the published 
information.  Therefore, the fascinating problem of interactions between
the tunneling defects should be left as an open question at this time.

We thank Eric Smith and Ch.  L. Spiel for their help and many
stimulating discussions and  Kris Poduska for the loan of the LR700 
resistance bridge.  We also thank J. Classen for sending us his 
preprint\cite{classen99}.  This work was supported by the National Science
Foundation, grant No.DMR-970972, and DMR9705295.

\begin{thebibliography}{10}

\bibitem{phillips87}
W.~A. Phillips, Rep. Prog. Phys. {\bf 50},  1657  (1987).

\bibitem{vancleve91}
J.~E. Van~Cleve, Ph.D. thesis, Cornell University, 1991, unpublished.

\bibitem{toppPRB99}
K. Topp, E. Thompson, and R.~O. Pohl, Phys. Rev. B {\bf 60},  898  (1999).

\bibitem{Nishiyama92}
H. Nishiyama, H. Akimoto, Y. Okuda, and H. Ishimoto, J. Low Temp. Phys. {\bf
  89},  727  (1992).

\bibitem{esquinazi92}
P. Esquinazi, R. Koenig, and F. Pobell, Z. Physik {\bf 87},  305  (1992).

\bibitem{classen94}
J. Classen {\it et~al.}, Ann. Physik {\bf 3},  315  (1994).

\bibitem{Rogge97}
S. Rogge, D. Natelson, B. Tigner, and D. Osheroff, Phys. Rev. B {\bf 55},
  11256  (1997).

\bibitem{classen99}
J. Classen, T. Burkert, E. Enss, S. Hunklinger, submitted to PRL.

\bibitem{enss97}
C. Enss and S. Hunklinger, Phys. Rev. Lett. {\bf 79},  2831  (1997).

\bibitem{burin98}
A. Burin, D. Natelson, D.and~Osheroff, and Y. Kagan,  in {\em Tunneling Systems
  in Amorphous and Crystalline Solids}, edited by P. Esquinazi (Springer
  Verlag, Berlin, 1998).

\bibitem{cahilljvc89}
D.~G. Cahill and J.~E. Van~Cleve, Rev. Sci. Instr. {\bf 60},  2706  (1989).

\bibitem{parpia85}
J. Parpia {\it et~al.}, Rev. Sci. Instr. {\bf 56},  437  (1985).

\bibitem{li86}
Q. Li {\it et~al.}, Cryogenics {\bf 26},  467  (1986).

\bibitem{Parpia83}
J. Parpia and T. Rhodes, Phys. Rev. Lett. {\bf 51},  805  (1983).

\bibitem{liu99}
X. Liu, E. Thompson, B.~E. White~Jr., and R.~O. Pohl, Phys. Rev. B {\bf 59},
  11767  (1999).

\bibitem{swartz89}
E.~T. Swartz and R.~O. Pohl, Rev. Mod. Phys. {\bf 61},  605  (1989).

\bibitem{cheeke76}
J.~D.~N. Cheeke, H. Ettinger, and B. Hebral, Can. J. Phys. {\bf 54},  1749
  (1976).

\bibitem{strehlow99physB}
P. Strehlow and M. Mei$\ss$ner, Physica B {\bf 263},  273  (1999).

\bibitem{strehlow99short}
P. Strehlow, {\it {et. al.}} submitted to PRL.

\end{thebibliography}

\end{multicols}{}
\end{document}